\documentclass[useAMS,usegraphicx,usenatbib]{mn2e}
\usepackage{times}

\begin{document}

\title[Different effects on constraints]{The effect of different observational data on the constraints of cosmological parameters}

\author[Y.G. Gong et al.]{Yungui Gong,$^{1,2,3}$\thanks{yggong@mail.hust.edu.cn} Qing Gao,$^1$\thanks{gaoqing01good@163.com} and Zong-Hong Zhu,$^4$
\thanks{zhuzh@bnu.edu.cn}\\
$^1$College of Mathematics and Physics, Chongqing University of Posts and Telecommunications,
Chongqing 400065, China\\
$^2$MOE Key Laboratory of Fundamental Quantities Measurement, School of Physics, Huazhong University of Science and Technology, Wuhan 430074, China\\
$^3$Institute of Theoretical Physics, Chinese Academy of Sciences, Beijing 100190, China\\
$^4$Department of Astronomy, Beijing Normal university,
Beijing 100875, China}
\maketitle

\begin{abstract}
The constraints on the $\Lambda$ cold dark matter ($\Lambda$CDM) model from type Ia supernova (SNe Ia) data alone and BAO data alone are similar,
so it is worthwhile
to compare their constraints on the property of dark energy.
We apply the three-year Supernova Legacy Survey (SNLS3) compilation of 472 SNe Ia data,
the baryon acoustic oscillation measurement of distance,
the cosmic microwave background radiation data from the seven-year {\it Wilkinson Microwave Anisotropy Probe} (WMAP7),
and the Hubble parameter data to study the effect of their different combinations on the fittings of cosmological parameters
in the modified holographic dark energy model and the Chevallier-Polarski-Linder model.
Neither BAO nor WMAP7 data alone give good constraint on the equation of state parameter of dark energy,
but both WMAP7 data and BAO data help SNe Ia data break the degeneracies
among the model parameters, hence tighten the constraint on the variation of equation of state parameter $w_a$,
and WMAP7 data do the job a little better.
Although BAO and WMAP7 data provide reasonably good constraints on $\Omega_m$ and $\Omega_k$, they are not able
to constrain the dynamics of dark energy. On the other hand, SNe Ia data do not provide good constraints on $\Omega_m$ and $\Omega_k$,
but they provide good constraint on the dynamics of dark energy, especially the variation of the equation of state parameter of dark energy,
so we need to combine SNe Ia with BAO and WMAP7 data to probe the property of dark energy. The addition of $H(z)$ data helps better constrain the geometry of the Universe $\Omega_k$ and the property of dark energy.
For the SNLS SNe Ia data, the nuisance parameters $\alpha$ and $\beta$
are consistent for all different combinations of the above data, and their impacts on the fittings of cosmological parameters are minimal.
By fitting the data to different models, $\Lambda$CDM model is still consistent with all the observational data.

\end{abstract}

%\pacs{98.80.Cq; 98.80.-k}
%\preprint{arXiv: 1110.6535}
%\preprint{CAS-KITPC/ITP-282}


\section{Introduction}
The accelerating expansion of the Universe was first discovered in 1998 by
the observations of Type Ia supernovae (SNe Ia) \citep{acc1,acc2}.
As more accurate data are available, it is possible to measure the acceleration and the dynamical mechanism behind
the acceleration. There are three different possibilities for the acceleration.
The first possibility is that a new
exotic form of matter with negative pressure, dubbed as dark energy drives the Universe to accelerate.
The cosmological constant is the simplest candidate of dark energy which is also consistent
with observations, but at odds with quantum field theory.
The second possibility is that general relativity is modified
at the cosmological scale, such as Dvali-Gabadadze-Porrati (DGP) model \citep{dgp}. The third possibility is that the
Universe is inhomogeneous. In this paper, we consider the possibility of dark energy only.

In the recent release of the measurements of the baryon acoustic oscillation (BAO) peaks at redshifts $z=0.44$, 0.6
and 0.73 in the galaxy correlation function of the final data set of the WiggleZ dark energy survey,
\cite{wigglez} used these three BAO data along with the BAO data at redshifts $z=0.2$ and 0.35 measured from the
distribution of galaxies \citep{wjp}
and the measurement of BAO at redshift $z=0.106$ from the six-degree Field Galaxy Survey (6dFGS) \citep{6dfgs}
to constrain $\Lambda$ cold dark matter ($\Lambda$CDM) model.
It was found that the constraints on $\Lambda$CDM model from the BAO data
are even better than those from Union2 SNe Ia data \citep{union2}.
\cite{wigglez} also found that the combination of BAO and the seven-year {\it Wilkinson Microwave
Anisotropy Probe} (WMAP7) data  gives
much better constraints on $\Lambda$CDM model than the combination of SNe Ia and WMAP7 data does.
This means that the updated BAO data \citep{wigglez} are
robust to constrain cosmological parameters.
If the constraints
on the property of dark energy from BAO data are much tighter than those from SNe Ia data, then
we just need to apply BAO data only for a faster fitting although SNe Ia data and BAO data are
complementary to each other. The redshifts of BAO data span from $z=0.106$ to $z=0.73$,
we may expect that BAO data catch the
dynamical property of dark energy, so it is necessary
to study the effects of different observational data and their combinations
on the constraints on the equation of state of dark energy.
In this paper, we use a simple dark energy model to test the robustness
of BAO data, and compare the constraints on the equation of state of dark energy from different data.

The question whether dark energy is just the cosmological constant remains to be answered. Recently,
there are lots of studies in determining whether $\Lambda$CDM model is consistent with observations
\citep{huang,star,cai,lampeitl,corray9,gong10a,gong10b,pan,gong11,li11}.
Through the reconstruction of $Om(z)=[E^2(z)-1]/[(1+z)^3-1]$ with the dimensionless Hubble parameter $E(z)=H(z)/H_0$ \citep{omz},
\cite{yu11} considered the tensions
between different data set by using Chevallier-Polarski-Linder (CPL) parametrization \citep{cpl1,cpl2} of the equation of state of dark energy. They found that a tension
between low redshift and high redshift data existed. \cite{cai11} used the figure of merit (FOM) proposed by the Dark
Energy Task Force \citep{detf} as a diagnostic to study the effectiveness of different combinations of data
on constraining $w_0$ and $w_a$ in CPL model.

In this paper, we first study the robustness of BAO data, then study the constraints on the equation of state of dark energy based
on different combinations of the following data:
the three-year Supernova Legacy Survey (SNLS3) sample of 472 SNe Ia data with systematic errors \citep{snls3};
the BAO measurements from the 6dFGS \citep{6dfgs}, the distribution of galaxies \citep{wjp} in the Sloan Digital Sky Survey (SDSS)  and the WiggleZ dark
energy survey \citep{wigglez}; the WMAP7 data \citep{wmap7};
and the Hubble parameter $H(z)$ data \citep{hz2,hz1}. In addition to studying the effects of
different observational data and their combinations on the constraints of cosmological parameters,
we also reconstruct the equation of state
of dark energy $w(z)$, the deceleration parameter $q(z)$ and $Om(z)$ by using these data sets.
On the other hand, the distance measurements from SNe, BAO and WMAP7 data
depend on $w(z)$ through double integrations, the process of double
integrations smoothes out the variation of $w(z)$.
Since the Hubble parameter $H(z)$ depends on $w(z)$ through a single integration,
the Hubble parameter $H(z)$ can detect the variation of $w(z)$ better than the distance scales
and needs to be applied to fit cosmological models.

The paper is organized as follows. In section 2, we present the SNLS3 SNe Ia data \citep{snls3},
the BAO data \citep{6dfgs,wigglez,wjp}, the WMAP7 data \citep{wmap7},
the $H(z)$ data \citep{hz2,hz1}, and
all the formulae related to these data.
In section 3, we present all the models and the fitting results,
and conclusions are drawn in section 4.

\section{Observational data}

\subsection{SNe Ia data}
The SNLS3 SNe Ia data consist of 123 low-redshift SNe Ia data with $z\la 0.1$
mainly from Calan/Tololo, CfAI, CfAII, CfAIII and CSP,
242 SNe Ia over the redshift range $0.08<z<1.06$ observed from the SNLS \citep{snls3},
93 intermediate-redshift SNe Ia data with $0.06\la z\la 0.4$ observed during the first season of Sloan Digital Sky
Survey (SDSS)-II supernova (SN) survey \citep{sdss2}, and 14 high-redshift SNe Ia data with $z\ga 0.8$
from {\it Hubble Space Telescope} \citep{hstdata}.
The SNLS3 SNe Ia data used the combination of SALT2 and SiFTO light-curve fitters \citep{snls3}.
To use the 472 SNLS3 SNe Ia data \citep{snls3},
we minimize
\begin{equation}
\label{chi}
\chi_{sn}^2(\mathbf{p},\alpha,\beta)=\sum_{i,j=1}^{472}(m_B-m_{mod})^{T}C^{-1}_{sn}(m_B-m_{mod}),
\end{equation}
where $m_B$ is the rest-frame peak B-band magnitude of a SN, the predicted magnitude of the SN
given a cosmological model is
$m_{mod}=5\log_{10}\mathcal{D}_L(z_{hel}, z_{cmb}, \mathbf{p})-\alpha (s-1)+\beta\mathcal{C}+\mathcal{M}_B$,
$z_{hel}$ and $z_{cmb}$ are the heliocentric and the CMB frame redshifts of the SN,
$s$ is the stretch given by the data, $\mathcal{C}$ is the color measure for the SN given by the data,
$\alpha$ and $\beta$ are nuisance parameters used for the SNLS3 data fitting, $\mathcal{M}_B$
is another nuisance parameter incorporating the absolute magnitude and Hubble constant and
it is marginalized over in the SNe data fitting process because of the arbitrary normalization of the magnitude,
$C_{sn}(z_i,z_j)$ is the covariant matrix which includes
both the systematical and statistical uncertainties for the SNe Ia data \citep{snls3}.
The correction on the dependence of the host-galaxy stellar mass is also included.
The Hubble-constant free luminosity distance $\mathcal{D}_L(z)$ is
\begin{equation}
\label{lum} \mathcal{D}_L(z)=H_0 d_L(z)=\frac{1+z}{\sqrt{|\Omega_{k}|}}\, S_k\!\left[\sqrt{|\Omega_{k}|}\int_0^z \frac{dx}{E(x)}\right],
\end{equation}
where the dimensionless Hubble parameter $E(z)=H(z)/H_0$
and $S_k(x)$ is defined as $x$, $\sin(x)$ or $\sinh(x)$ for $k=0$, +1, or -1, respectively.
For the fitting to the SNLS3 data, we need to add two more nuisance parameters $\alpha$ and $\beta$
in addition to the model parameters $\mathbf{p}$ and the nuisance parameter $\mathcal{M}_B$.

\subsection{BAO data}

For the BAO data, we use the measurements from the 6dFGS \citep{6dfgs}, the distribution of galaxies \citep{wjp} in the SDSS
and the WiggleZ dark energy survey \citep{wigglez}.
\cite{wjp} measured the distance ratio,
\begin{equation}
\label{dz} d_{z}= \frac{r_{s}(z_{d})}{D_{V}(z)}
\end{equation}
at two redshifts $z=0.2$ and $z=0.35$ by fitting to the power spectra of luminous red galaxies and main-sample galaxies in the SDSS.
Here the effective distance is
\begin{equation}
\label{dvdef}
D_V(z)=\left[\frac{d_L^2(z)}{(1+z)^2}\frac{z}{H(z)}\right]^{1/3},
\end{equation}
the drag redshift $z_d$ is \citep{dw},
\begin{equation}
\label{zdfiteq} z_d=\frac{1291(\Omega_m
h^2)^{0.251}}{1+0.659(\Omega_m h^2)^{0.828}}[1+b_1(\Omega_b h^2)^{b_2}],
\end{equation}
where
\begin{eqnarray}
\label{b1eq} b_1&=&0.313(\Omega_m h^2)^{-0.419}[1+0.607(\Omega_m
h^2)^{0.674}],  \\
b_2&=&0.238(\Omega_m h^2)^{0.223},
\end{eqnarray}
the comoving sound horizon is
\begin{equation}
\label{rshordef}
r_s(z)=\int_z^\infty \frac{c_s(x)dx}{H(x)},
\end{equation}
the sound speed $c_s(z)=1/\sqrt{3[1+\bar{R_b}/(1+z)}]$, and
$\bar{R_b}=3\Omega_b h^2/(4\times2.469\times10^{-5})$.
\cite{6dfgs} derived that $d_{0.106}^{obs}=0.336\pm 0.015$
from the 6dFGS measurements.
The WiggleZ dark energy survey measured the acoustic parameter
\begin{equation}
\label{apardef}
A(z)=\frac{D_V(z)\sqrt{\Omega_m H_0^2}}{z},
\end{equation}
at three redshifts $z=0.44$, $z=0.6$ and $z=0.73$.
To use the BAO data, we minimize
\begin{eqnarray}
\label{baochi2}
\chi^2_{Bao}(\mathbf{p},\Omega_b h^2, h)&=&\sum_{i,j=1}^{2}\Delta d_i
C_{dz}^{-1}(d_i,d_j)\Delta d_j\nonumber\\
&&+\frac{(d_{0.106}-0.336)^2}{0.015^2}\nonumber\\
&&+\sum_{i,j=1}^{3}\Delta A_i C_{A}^{-1}(A_i,A_j)\Delta A_j,
\end{eqnarray}
where $d_i=(d_{z=0.2},d_{z=0.35})$, $\Delta d_i=d_i-d_i^{obs}$ and
the covariance matrix $C_{dz}(d_i,d_j)$ for $d_z$ at $z=(0.2,0.35)$
is taken from equation (5) in \cite{wjp};
$A_i=(A(0.44),A(0.6),A(0.73))$, $\Delta A_i=A(z_i)-A(z_i)^{obs}$
and the covariance matrix $C_A(A_i,A_j)$ for the data points $A(z)$ at
$z=(0.44,0.6,0.73)$ is taken from table 2 in \cite{wigglez}.
The BAO data listed in Table \ref{baotab} span the redshift regions $0.106-0.73$ in which dark energy is supposed to dominate cosmic expansion,
so we naively expect that the BAO data can catch the dynamics of dark energy.
Besides the model parameters $\mathbf{p}$, we need
to add two more nuisance parameters $\Omega_b h^2$ and $\Omega_m h^2$ when we use the BAO data from 6dfGS \citep{6dfgs} and SDSS \citep{wjp}.

\begin{table}
\begin{center}
\caption{The BAO distance data from the 6dFGS \citep{6dfgs}, SDSS \citep{wjp} and WiggleZ surveys \citep{wigglez}. \label{baotab}}
\begin{tabular}{cccc}
Data & $z$ & $d_z$ & $A(z)$ \\ \hline
6dFGS & 0.106 & $0.336\pm 0.015$ & \\
SDSS & 0.2 & $0.1905\pm 0.0061$ & \\
SDSS & 0.35 & $0.1097\pm 0.0036$ & \\
WiggleZ & 0.44 & & $0.474\pm 0.034$ \\
WiggleZ & 0.6 &  & $0.442\pm 0.020$ \\
WiggleZ & 0.73 & & $0.424\pm 0.021$ \\ \hline
\end{tabular}
\end{center}
\end{table}

\subsection{WMAP7 data}

For the WMAP7 data, we use the measurements of the three derived quantities: the shift parameter
$R(z^{*})$ and the acoustic index $l_A(z^{*})$ at the recombination redshift $z^{*}$.
The shift parameter $R$ is
expressed as
\begin{equation}
\label{shift}
R(z^{*})=\frac{\sqrt{\Omega_{m}}\mathcal{D}_L(z^*)}{1+z^*}.
\end{equation}
 The acoustic index $l_A$ is
\begin{equation}
\label{ladefeq} l_A(z^{*})=\frac{{\rm \pi}
d_L(z^{*})}{(1+z^{*})r_s(z^{*})},
\end{equation}
and the recombination redshift $z^{*}$ is fitted by  \citep{hs},
\begin{eqnarray}
\label{zstareq} z^{*}=1048[1+0.00124(\Omega_b
h^2)^{-0.738}][1+g_1(\Omega_m h^2)^{g_2}],
\end{eqnarray}
\begin{equation}
g_1=\frac{0.0783(\Omega_b h^2)^{-0.238}}{1+39.5(\Omega_b
h^2)^{0.763}},\quad g_2=\frac{0.560}{1+21.1(\Omega_b h^2)^{1.81}}.
\end{equation}
In particular, we minimize
\begin{equation}
\label{cmbchi} \chi^2_{CMB}(\mathbf{p},\Omega_b h^2, h)=\sum_{i,j=1}^{3}\Delta x_i C_{CMB}^{-1}(x_i,x_j)\Delta x_j,
\end{equation}
where the three parameters $x_i=[R(z^{*}),\ l_A(z^{*}),\ z^{*}]$, $\Delta
x_i=x_i-x_i^{obs}$ and the covariance matrix $C_{CMB}(x_i,x_j)$
for the three parameters taken from table 10 in \cite{wmap7} are listed in Table \ref{wmap7tab}.
We also need to add the nuisance parameters $\Omega_b h^2$ and $\Omega_m h^2$ to the parameter space when we fit the WMAP7 data.
\begin{table}
\begin{center}
 \caption{The maximum likelihood (ML) and the inverse covariance matrix for the three distance priors \citep{wmap7}. \label{wmap7tab}}
\begin{tabular}{cccc|c}
\hline
$x_i$ & $l_A$ & $R$ & $z^*$ & ML  \\ \hline
$l_A$ & 2.305 & 29.698 & -1.333 & 302.09 \\
$R$ &  & 6825.270 & -113.180& 1.725 \\
$z^*$ &  &  & 3.414 & 1091.3 \\ \hline
\end{tabular}
\end{center}
\end{table}

\subsection{Hubble data}

The distances measured by the SNe, BAO and WMAP7 data depend on
the double integrations of the equation of state parameter $w(z)$, the process of double
integrations smoothes out the variation of equation of state parameter $w(z)$ of dark energy.
However, the Hubble parameter $H(z)$ depends on one integration of $w(z)$.
Therefore, the Hubble parameter $H(z)$ can detect the variation of $w(z)$ better than the distance scales do.
Furthermore, it was found that $w(z)$ at high redshifts will be better
constrained with the addition of $H(z)$ data \citep{gong10a}.
So we also use the $H(z)$ data at 11 different redshifts obtained from the differential ages of
passively evolving galaxies  \citep{Simon:2004tf,hz1}, and three more Hubble parameter data
at redshifts $z=0.24$, $z=0.34$ and $z=0.43$, determined by taking the BAO scale as a standarad ruler in the radial driection \citep{hz2}.
The $H(z)$ data span out to the redshift regions $z=1.75$ and is shown in Table \ref{hzdatatab}. So we add these $H(z)$ data to $\chi^2$,
\begin{equation}
\label{hzchi}
\chi^2_H(\mathbf{p}, h)=\sum_{i=1}^{14}\frac{[H(z_i)-H_{obs}(z_i)]^2}{\sigma_{hi}^2},
\end{equation}
where $\sigma_{hi}$ is the $1\sigma$ uncertainty of $H(z)$.

\begin{table}
\begin{center}
\caption{The 14 $H(z)$ data from sources a \citep{hz1}, b \citep{Simon:2004tf} and c \citep{hz2}.  \label{hzdatatab}}
\begin{tabular}{ccc}
\hline
z & $H(z)$ [km/sec/Mpc] & sources  \\ \hline
0.1 & $69\pm 12$ & a\\
0.17 & $83\pm 8$ & a \\
0.27 & $77\pm 14$ & a \\
0.4 & $95\pm 17$ & a \\
0.48 & $97\pm 62$ & a \\
0.88 & $90\pm 40$ & a \\
0.9 & $117\pm 23$ & a \\
1.3 & $168\pm 17$ & b \\
1.43 & $177\pm 18$ & b \\
1.53 & $140\pm 14$ & b \\
1.75 & $202\pm 40$ & b \\
0.24 & $76.69\pm 2.32$ & c \\
0.34 & $83.8\pm 2.96$ & c \\
0.43 & $86.45\pm 3.27$ & c \\
\hline
\end{tabular}
\end{center}
\end{table}

\subsection{Fitting method}

Basically,
the model parameters $\mathbf{p}$ are determined by minimizing
\begin{equation}
\label{chi2min}
\chi^2=\chi^2_{sn}+\chi^2_{Bao}+\chi^2_{CMB}+\chi^2_H.
\end{equation}
The likelihood for the parameters $\mathbf{p}$ in the
model and the nuisance parameters is
computed using the Monte Carlo Markov Chain (MCMC) method.
The MCMC method
randomly chooses values for the above parameters, evaluates $\chi^2$
and determines whether to accept or reject the set of parameters
using the Metropolis-Hastings algorithm. The set of parameters that
are accepted to the chain forms a new starting point for the next
process, and the process is repeated for a sufficient number of
steps until the required convergence is reached. Our MCMC code is
based on the publicly available package {\sc cosmomc} \citep{cosmomc,gong08}.
When SNe Ia data are used, we also need to fit the two nuisance parameters
$\alpha$ and $\beta$, when BAO or WMAP7 data are used, we need to fit
the two nuisance parameters $\Omega_b h^2$ and $h=H_0/100$. So
when we combine SNe Ia with BAO or WMAP7 data, we need to fit four
nuisance parameters $(\alpha,\beta,\Omega_b h^2,h)$.

After fitting the observational data to different dark energy models, we apply the $Om$ diagnostic \citep{omz}
to detect the deviation from the $\Lambda$CDM model.
For the $\Lambda$CDM model with $\Omega_k=0$ (we call flat model afterwards),
$Om(z)=\Omega_m$ is a constant which is independent of the value of $\Omega_m$.
Because of this property, $Om$ diagnostic is less sensitive to observational errors
than the equation of state parameter $w(z)$ does.
On the other hand, the bigger the value of $Om(z)$, the bigger the value of $w(z)$,
so the behavior of $Om(z)$ catches the dynamical property of $w(z)$.

We also apply the FOM as a diagnostic tool to compare the effectiveness of different combinations of observational data
on constraining the equation of state parameters $w_0$ and $w_a$ in CPL model. FOM is defined as the
the reciprocal of the area of the error ellipse
enclosing the 95\% confidence limit in the $w_0$-$w_a$ plane, it is proportional to $[{\rm det}C_w(w_0,w_a)]^{-1/2}$,
here $C_w(w_0,w_a)$ is the correlation matrix of $w_0$ and $w_a$.

\section{Cosmological fitting results}
\subsection{$\Lambda$CDM model}

\begin{figure*}
\includegraphics[width=0.95\textwidth]{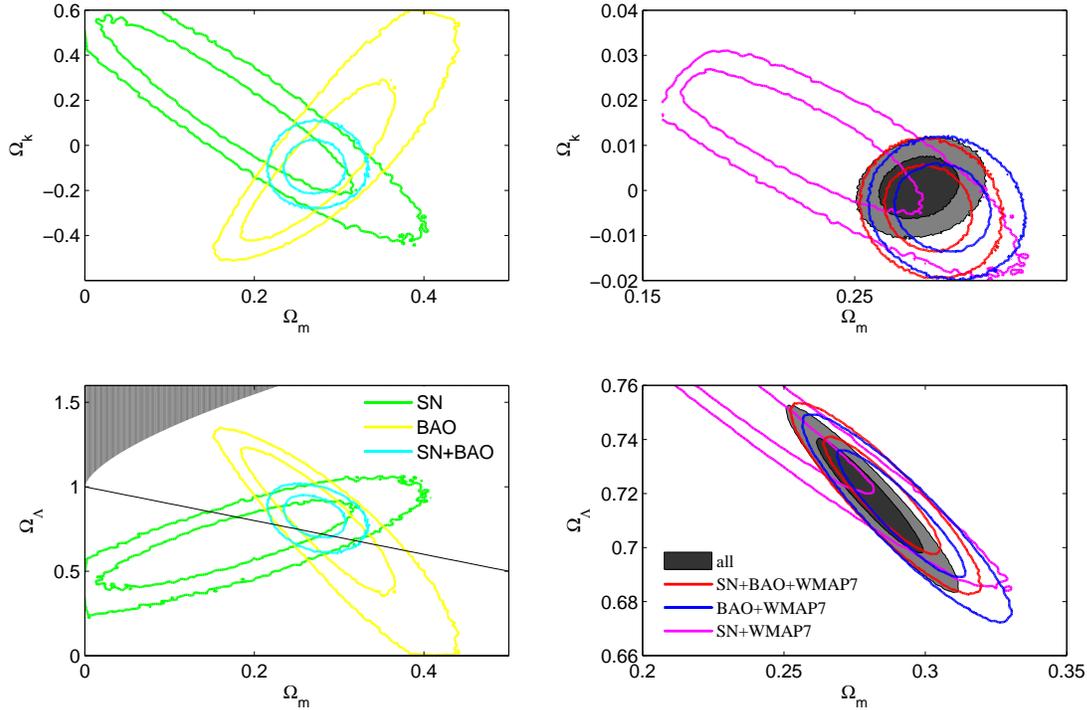}
\caption{The marginalized $1\sigma$ and $2\sigma$ contour plots of $\Omega_m$ and $\Omega_k$,
and $\Omega_m$ and $\Omega_\Lambda$
for the curved $\Lambda$CDM model.
In the left-hand panels, we use the following data: SNe data alone, BAO data alone, and the combination of SNe and BAO data.
In the right-hand panels,
we use the following data: the combination of SNe and WMAP7 data, the combination of BAO and WMAP7 data,
the combination of SNe, BAO and WMAP7, and all the data combined.
The green lines label the constraints from SNe Ia data only, the yellow
lines label the constraints from BAO data only, the cyan lines label the constraints from the combination of SNe Ia and BAO data,
the magenta lines label the constraints from the combination of SNe Ia and WMAP7 data,
the blue lines label the constraints from the combination of WMAP7 and BAO data,
the red lines label the constraints from the combination of SNe Ia, BAO and WMAP7 data,
and the shaded regions label the constraints from the combination of all the observational data described in section 2.
The black solid line the lower-left panel denotes the flat $\Lambda$CDM model.}
\label{lcdmcont}
\end{figure*}

We first review the effects of different combinations of observational data on
the $\Lambda$CDM model with non-zero $\Omega_k$ (we call it curved model afterward).
The $\Omega_m$-$\Omega_\Lambda$ contour from applying only the SNLS3 SNe data was shown in figure 8 in \cite{snls3}.
By combining the SNLS3 SNe and the WMAP7 data, \cite{sullivan11} obtained the constraints on $\Omega_m$ and $\Omega_k$,
and the contours were shown in figure 5 of their paper. From these results,
we see that SNLS3 SNe data alone do not provide tight constraints on $\Omega_m$ and $\Omega_k$. With
the addition of WMAP7 data, the constraint on $\Omega_k$ becomes much tighter, so the constraint on $\Omega_m$ becomes tighter.
Therefore, WMAP7 data can be used to tighten the constraint on the geometry of the Universe \citep{wigglez}.
By applying the BAO data only with the assumption that $\Omega_b h^2=0.02227$,
\cite{wigglez} found the constraints on $\Omega_m$ and $\Omega_k$.
Comparing the constraints from SNLS3 or Union2 SNe Ia data alone with those from BAO data alone,
we see that the constraints are similar,
and the constraint on $\Omega_m$ from BAO data alone is even much better than that from SNe Ia data alone.
\cite{wigglez} also compared the constraint on the curved $\Lambda$CDM model from
the combinations of WMAP7 with BAO and Union2 SNe Ia data,
and they found that $\Omega_m$-$\Omega_k$ contours from the combination of BAO and WMAP7 data are smaller
than those from the combination of Union2 SNe Ia and WMAP7 data.
Moreover, the constraints from the combination of Union2 SNe Ia, BAO and WMAP7 data are similar to
those from the combination of BAO and WMAP7 data.
These results show that BAO data mainly tighten the constraint on $\Omega_m$ and WMAP7 data mainly tighten
the constraint on $\Omega_k$, while current SNe Ia data still give large $\Omega_m$-$\Omega_k$ contours,
so the addition of SNe data to the combination of BAO and WMAP7 data has little effect in improving
the constraints of $\Omega_m$ and $\Omega_k$. For comparison,
we show all the constraints in Fig. \ref{lcdmcont}.
Note that we set the nuisance parameters $\Omega_b h^2$ and $h$ as free parameters in fitting BAO and WMAP7 data,
so we fit the four parameters $(\Omega_m,\ \Omega_k,\ \Omega_bh^2,\ h)$ when BAO or WMAP7 data are used,
the four parameters $(\Omega_m,\ \Omega_k,\ \alpha,\ \beta)$ when SNe data are used,
and six parameters $(\Omega_m,\ \Omega_k,\ \Omega_bh^2,\ h,\ \alpha,\ \beta)$ when we combine SNe with BAO or WMAP7 data.
In Fig. \ref{lcdmcont}, we show the constraints on the curved $\Lambda$CDM model
from SNLS3 SNe Ia data alone (the green lines), BAO data alone (the yellow line),
the combination of SNLS3 SNe Ia and BAO data (the cyan lines), the combination of SNLS3 SNe Ia and WMAP7 data (the magenta lines),
the combination of BAO and WMAP7 data (the blue lines), the combination of SNLS3 SNe Ia, BAO and WMAP7 data (the red lines)
 and the combination of all observational data (the shaded regions).
In the left-hand panels, we use the following data: SNe data alone, BAO data alone, and the combination of SNe and BAO data.
In the right-hand panels,
we use the following data: the combination of SNe and WMAP7 data, the combination of BAO and WMAP7 data, the combination of SNe, BAO and WMAP7 data
and all the data combined.
These results are summarized in Table \ref{table1}.

Although the binned SNe data measure the distance-redshift relation at $z<0.8$ with three to four
times higher accuracy than the BAO data \citep{wigglez}, the constraints from SNe and BAO data are similar
and BAO data constrain $\Omega_m$ even better.  As it is well known,
the directions of degeneracy between $\Omega_m$ and $\Omega_\Lambda$ from
SNe and BAO data are different,
so the combination of these two data sets can improve the accuracy of the constraints much better.
However, the uncertainties on $\Omega_m$ and $\Omega_k$ constrained from the combination
of BAO and WMAP7 data become much smaller than those from SNe data alone, so the addition of SNe data
to the combination of BAO and WMAP7 data has little effect even though the degeneracy directions are different.
The addition of $H(z)$ data further reduces the uncertainty of $\Omega_k$ and moves the best-fitting value of $\Omega_k$ towards zero.
We also find that the uncertainties of the nuisance parameters $\alpha$ and $\beta$ are around $0.1$, and they
are all consistent at $1\sigma$ level for different fittings.

\begin{table}
\begin{center}
 \caption{The marginalized $1\sigma$
errors for $\Omega_m$ and $\Omega_k$ in curved $\Lambda$CDM model constrained by different observational data\label{table1}}
\begin{tabular}{ccc}
\hline
Data & $\Omega_{m}$ & $\Omega_{k}$  \\ \hline
SNe Ia & $0.17^{+0.1}_{-0.09}$ & $0.15 \pm 0.25$  \\ \hline
BAO & $0.26^{+0.09}_{-0.03}$ & $-0.16^{+0.38}_{-0.11}$  \\ \hline
SNe+BAO & $0.27\pm 0.02$ & $-0.11_{-0.07}^{+0.09}$ \\ \hline
SNe+WMAP7 & $0.22^{+0.05}_{-0.03}$ & $0.01\pm 0.01$  \\ \hline
BAO+WMAP7 & $0.29^{+0.02}_{-0.01}$ & $-0.004_{-0.006}^{+0.007}$  \\ \hline
SNe+BAO+WMAP7 & $0.28\pm 0.01 $ & $-0.004^{+0.006}_{-0.007}$  \\ \hline
All & $0.28^{+0.02}_{-0.01}$ & $0.0006_{-0.0045}^{+0.0046}$  \\ \hline
\end{tabular}
\end{center}
\end{table}

\begin{figure}
\includegraphics[width=0.45\textwidth]{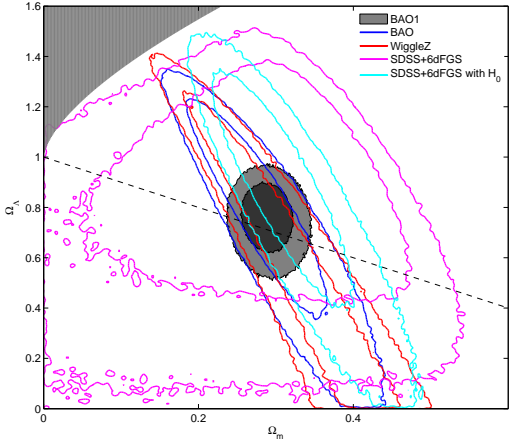}
\caption{The marginalized $1\sigma$ and $2\sigma$ contour plots of $\Omega_m$ and $\Omega_\Lambda$
for the curved $\Lambda$CDM model by using different BAO data.
The magenta contours are for the the combination of SDSS and 6dFGS BAO data by assuming $\Omega_b h^2=0.02227$,
the cyan contours are for the the combination of SDSS and 6dFGS BAO data by assuming $\Omega_b h^2=0.02227$ and $h=0.738$,
the red contours are for the WiggleZ BAO data, the blue contours are for the combination of 6dFGS, SDSS and WiggleZ BAO data with free $\Omega_b h^2$ and $h$,
and the shaded contours are for the BAO1 data with free $\Omega_b h^2$ and $h$.
The dashed black line denotes the flat $\Lambda$CDM model.}
\label{baoclmcont}
\end{figure}

By using the measurement of $d_z$ at $z=0.275$ from \cite{wjp}, the $\Omega_m$-$\Omega_\Lambda$ contours were plotted in
figure 5 in \cite{wjp} and figure 10 in \cite{union2}. Comparing those plots \citep{wjp,union2} with ours in Fig. \ref{lcdmcont},
it is clear that the updated BAO data \citep{wigglez} greatly improve the constraints on $\Omega_m$ and $\Omega_\Lambda$,
and the ability of constraining the curved $\Lambda$CDM model by current BAO data \citep{wigglez} is even better than that by SNe Ia data.
Therefore, it is interesting to study the ability of current BAO data on constraining the dynamical behavior of dark energy.
However, we need to understand what causes the improvement.
We would like to see wether it is due to more data or larger redshift regions the data spanned.
To do that, we compare the constraints on the curved $\Lambda$CDM model from individual BAO data.
For the BAO data from SDSS \citep{wjp} and 6dFGS \citep{6dfgs}, there
are two nuisance parameters $\Omega_b h^2$ and $h$ in addition to the
two model parameters $\Omega_m$ and $\Omega_k$, so we need to impose priors
to get some reasonable results. Following \cite{wjp}, we fix $\Omega_b h^2=0.02227$
and consider both cases with $h=0.738$ \citep{Riess:2011yx} and free $h$.
The results are shown in Fig. \ref{baoclmcont}. When $h$ is a free parameter,
$\Omega_m$ is not well constrained by the combination of 6dFGS and SDSS data (the magenta contours).
However, when we take $h=0.738$ (the cyan contours), the constraint on $\Omega_m$ is greatly improved.
We also check the case with the prior $H_0=73.8\pm 2.4$ \citep{Riess:2011yx},
and the situation is similar although the improvement over free $H_0$ is smaller.
For the BAO data from WiggleZ \citep{wigglez},
we only have two model parameters $\Omega_m$ and $\Omega_k$ and the results are shown by the red contours in Fig. \ref{baoclmcont}.
We see that the constraints from WiggleZ data (the red contours)
are similar to those from the combination of 6dFGS and SDSS  with nuisance parameters fixed (the cyan contours)
and the combination of 6dFGS, SDSS and WiggleZ data with free nuisance parameters (the blue contours),
and these results are consistent with each other.
We also check the constraints from one of the three $A(z)$ data presented in the WiggleZ,
and we find that the result is similar. The result that the parameter $A(z)$ gives tighter constraint on $\Omega_m$ was also found in \cite{gong10a,gong10b}.

Recently, \cite{ngbusca} reported the detection of BAO in the Ly$\alpha$ forest of high-redshift quasars from the Baryon Oscillation Spectroscopic Survey.
They found that $H(z=2.3)r_s(z_d)/(1+z)=(1.036\pm 0.036)\times 10^4$ km/s which gives the radial BAO data $\Delta z(z)=H(z)r_s(z_d)/c=0.11404\pm 0.00396$
at the redshift $z=2.3$. In order to distinguish different BAO data, we refer the combination of 6dFGS, SDSS and WiggleZ BAO data as BAO. When
the radial BAO data at $z=2.3$ is added to the combination of 6dFGS, SDSS and WiggleZ BAO data, we call the data BAO1. With the addition of the radial BAO data
at high redshift, we expect better constraints on the cosmological parameters. The $1\sigma$ and $2\sigma$ contours
of $\Omega_m$ and $\Omega_k$ from BAO1 data are shown by the shaded regions in Fig. \ref{baoclmcont}.
We see that with the addition of BAO measurement at high redshift, the constraint on the curved $\Lambda$CDM model was greatly
improved, and the constraint from BAO1 is even much better than that from SNe Ia data.
These results suggest that more data points and more measurements at high redshift can help improve the results.

\subsection{Modified holographic dark energy model}

\begin{figure*}
\includegraphics[width=0.95\textwidth]{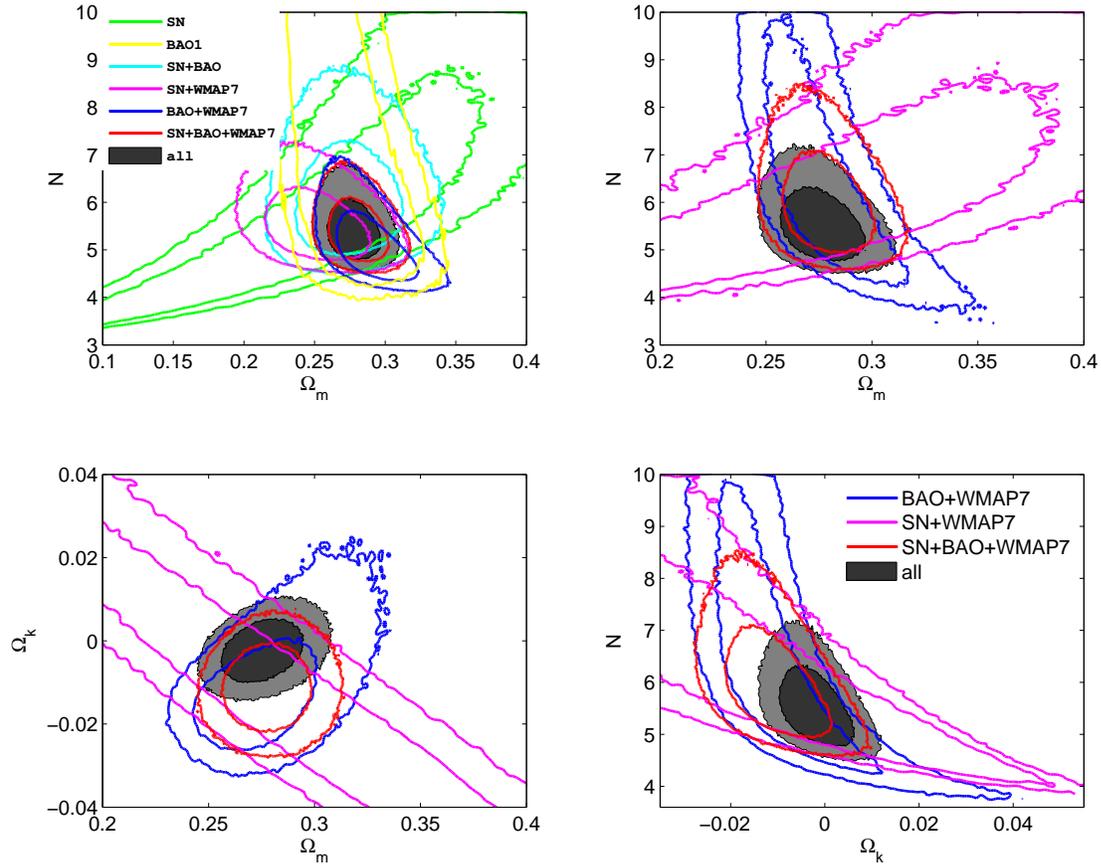}
\caption{The marginalized $1\sigma$ and $2\sigma$ contour plots for MHDE model.
The figures from the upper left to lower right
are labeled as (a)-(d), respectively. (a) is for the flat MHDE model and (b)-(d) are for the curved MHDE model.
In (a), we use the following data: SNe data alone, the combination of SNe and BAO data,
the combination of SNe and WMAP7 data, the combination of BAO and WMAP7 data, the combination of SNe, BAO and WMAP7, and all the data combined.
In (b)-(d), we use the following data:
the combination of SNe and WMAP7 data, the combination of BAO and WMAP7 data,
the combination of SNe, BAO and WMAP7, and all the data combined.
The green lines label the constraints from SNe Ia data only,
the yellow lines label the constraints from BAO1 data only,
the cyan lines label the constraints from the combination of SNe Ia and BAO data,
the magenta lines label the constraints from the combination of SNe Ia and WMAP7 data,
the blue lines label the constraints from the combination of WMAP7 and BAO data,
the red lines label the constraints from the combination of SNe Ia, BAO and WMAP7 data,
and the shaded regions label the constraints from the combination of all the observational data.}
\label{figure5}
\end{figure*}

Apart from the simple $\Lambda$CDM model, we consider a specific dark energy model which just has one more parameter
than $\Lambda$CDM model in this subsection.
Applying the relationship between the mass and the horizon of a Schwarzschild black hole in higher dimensions
and holographic principle, a modified holographic dark energy model (MHDE) with Hubble horizon as the ultraviolet cutoff
was proposed in \cite{gongli10}. Both the DGP model \citep{dgp} and $\Lambda$CDM model are special cases of this model,
note that here we consider DGP model as an effective dark energy model instead of a model which modifies gravity.
In this model, Friedmann equation is  \citep{gongli10,dvaliturner}
\begin{eqnarray}
\label{dgpez}
&E^2(z)-(1-\Omega_m-\Omega_k-\Omega_r)E^{5-N}(z) \nonumber\\
&=\Omega_{k}(1+z)^2+\Omega_{m}(1+z)^3+\Omega_r (1+z)^4,
\end{eqnarray}
where $N$ is the spatial dimension. So we recover the DGP model if $N=4$ and $\Lambda$CDM model if $N=5$.
This model has one more parameter than $\Lambda$CDM model, i.e., there are three model
parameters $\mathbf{p}=(\Omega_{m},\ \Omega_{k},  N)$ in this model
in addition to the four nuisance parameter ($\alpha$, $\beta$, $\Omega_b h^2$, $h$) to be fitted.
For the flat MHDE model, $\Omega_k=0$, we have only two model parameters $\Omega_m$ and $N$ and
we consider the constraints from SNe Ia data, the combination of SNe Ia and BAO data,
the combination of SNe Ia and WMAP7 data, the combination of BAO and WMAP7 data, the combination of SNe Ia,
BAO and WMAP7 data, and the combinations of all the observational data, the contours of
$\Omega_m$-$N$ are shown in Fig. \ref{figure5}(a), and the $1\sigma$ constraints are summarized in Table \ref{table3}.
The constraints on $N$ from BAO data alone are not good, so the results are not shown. Since BAO1 data alone gave
much better constraint on the $\Lambda$CDM model, we also apply BAO1 data to the flat MHDE model and the result
is shown in Fig. \ref{figure5}(a) by yellow contours. The constraints on $N$ from BAO1 data alone are not good either,
so the additional BAO measurement at $z=2.3$ does not seem to help improve the constraint on the dynamics of dark energy,
we mainly consider the effect of BAO data in the rest of the paper.
In the upper left panel in Fig. \ref{figure5}, we use the following data: SNe data alone, BAO1 data alone, the combination of SNe and BAO data,
the combination of SNe and WMAP7 data, the combination of BAO and WMAP7 data, and all the data combined.

Although BAO and BAO1 alone do not provide good constraints on $N$, the addition of BAO to SNe Ia data greatly improves the constraint on $\Omega_m$,
therefore improves the constraint on $N$.
The effect of the WMAP7 data is similar
except that the best-fitting values of the parameters $\Omega_m$ and $N$ become smaller. The effect of the combination
of BAO and WMAP7 data is similar to that of the combination of SNe Ia and WMAP7 data except that the best-fitting value
of $\Omega_m$ becomes bigger and the best-fitting value of $N$ becomes smaller. When we combine SNe Ia, BAO and WMAP7 data
or all the observational data, the results are similar, so the addition of $H(z)$ has little effect.
For all the combinations, $N\ga 5$ at $1\sigma$ level.
The SNLS3 SNe Ia data fitting parameters $\alpha$ and $\beta$ are also consistent for different data combinations.
These results show that both WMAP7 and BAO data help SNe data greatly improve the constraint on $\Omega_m$,
therefore improve the constraint on $N$. The degeneracy between $\Omega_m$ and $N$ obtained from BAO data
is different from that from WMAP7 data.

\begin{table}
\begin{center}
 \caption{The marginalized $1\sigma$ constraints on MHDE model by different observational data. The top
 six rows are for the flat MHDE model and the bottom four rows are for the curved MHDE model.
 \label{table3}}
\begin{tabular}{cccc}
\hline
Data & $\Omega_{m}$& $\Omega_{k}$& $N$  \\ \hline
SNe & $0.26\pm 0.09$ & & $6.1_{-1.8}^{+2.0}$ \\ \hline
SNe+BAO & $0.28^{+0.02}_{-0.03}$ & & $5.8^{+1.2}_{-0.5}$  \\ \hline
SNe+WMAP7 & $0.25_{-0.02}^{+0.03}$ & & $5.3_{-0.3}^{+0.8}$  \\ \hline
BAO+WMAP7 & $0.29\pm 0.02$ & & $4.9^{+0.8}_{-0.3}$ \\ \hline
SNe+BAO+WMAP7 & $0.28 \pm 0.01$ & & $5.3^{+0.6}_{-0.3}$ \\ \hline
All & $0.28\pm 0.01$ & & $5.3_{-0.2}^{+0.6}$ \\ \hline
\hline
BAO+WMAP7 & $0.28\pm 0.02$ & $-0.01\pm 0.01$ & $6.0_{-1.0}^{+2.5}$ \\ \hline
SNe+WMAP7 & $0.29_{-0.08}^{+0.07}$ & $-0.02\pm 0.03$ & $6.5_{-1.5}^{+1.7}$ \\ \hline
SNe+BAO+WMAP7 & $0.28^{+0.02}_{-0.01}$ & $-0.011\pm 0.007$ & $5.8_{-0.5}^{+1.0}$ \\ \hline
All & $0.28\pm 0.01$ & $-0.003\pm 0.005$ & $5.5^{+0.6}_{-0.4}$  \\ \hline
\end{tabular}
\end{center}
\end{table}

For the curved case, $\Omega_k\neq 0$, we have three model parameters.
From the results of $\Lambda$CDM model, we see that the constraints on $\Omega_k$ from either
SNe or BAO data alone are not good, and WMAP7 data help tighten the constraint on $\Omega_k$.
So in the curved model, we use WMAP7 data as priors.
Therefore we only consider the combinations of SNe and/or BAO data with WMAP7 data for the curved model in the rest of this paper.
The contours of $\Omega_m$-$N$, $\Omega_m$-$\Omega_k$,
and $\Omega_k$-$N$ are shown in Figs. \ref{figure5}(b)-(d), and the $1\sigma$ constraints are summarized in Table \ref{table3}.
In Figs. \ref{figure5}(b)-(d), we use the following data:
the combination of SNe and WMAP7 data, the combination of BAO and WMAP7 data,
the combination of SNe, BAO and WMAP7, and all the data combined.
For the constraints on $\Omega_m$
and $\Omega_k$, we see that the combination of BAO and WMAP7 data does better than the combination of SNe and WMAP7 data.
However, for the constraint on $N$, both combinations get similar results.
The degeneracies among the model parameters from the combination of SNe and BAO data and from the combination of BAO and WMAP7 data
are different, so when we combine SNe, BAO and WMAP7 data,
the constraints on the model parameters $\Omega_m$, $\Omega_k$ and $N$ are further improved. With the addition of $H(z)$ data,
the best-fitting value of $\Omega_k$ is moved toward zero and the upper limit of $N$ are reduced a little further.
The results also show that observational data favour $\Lambda$CDM model more than DGP model
since larger value of $N$ is favored. The SNLS3 SNe Ia data
fitting parameters $\alpha$ and $\beta$ are consistent for different data combinations.

\subsection{CPL parametrization}

\begin{figure*}
\includegraphics[width=0.95\textwidth]{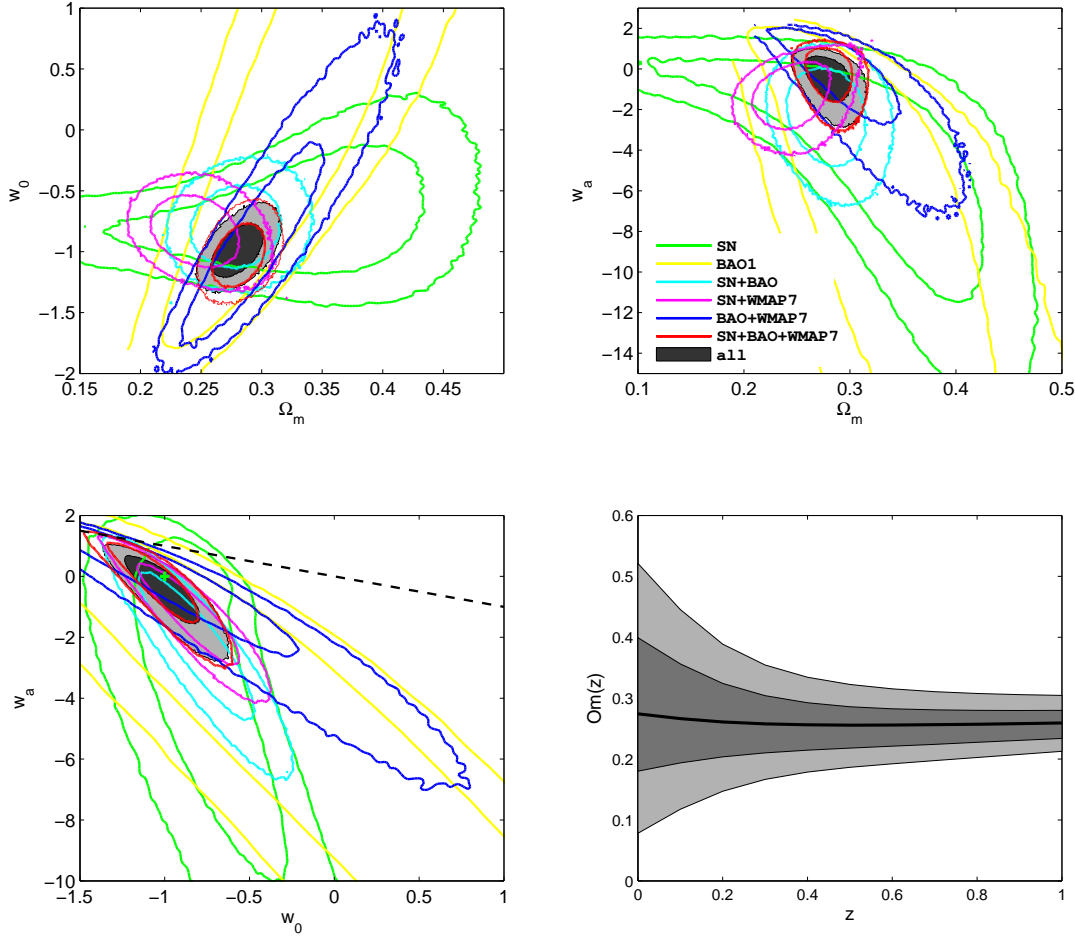}
\caption{The marginalized $1\sigma$ and $2\sigma$ contour plots of $\Omega_m$-$w_0$, $\Omega_m$-$w_a$ and $w_0$-$w_a$
for the flat CPL model. The green lines label the constraints from SNe Ia data only,
the yellow lines label the constraints from BAO1 data only,
the cyan lines label the constraints from the combination of SNe Ia and BAO data,
the magenta lines label the constraints from the combination of SNe Ia and WMAP7 data,
the blue lines label the constraints from the combination of WMAP7 and BAO data,
the red lines label the constraints from the combination of SNe Ia, BAO and WMAP7 data,
and the shaded regions label the constraints from the combination of all the observational data.
The dashed line in
the $w_0$-$w_a$ contour denotes the condition $w_0+w_a=0$, and the + sign denotes the point corresponding to the $\Lambda$CDM model.
In the lower right panel, we reconstruct $Om(z)$ by using the
constraints from the combination of all data for flat CPL model, the solid line is obtained
by using the best-fitting values of $\Omega_m$, $w_0$ and $w_a$.}
\label{figure3}
\end{figure*}

\begin{figure}
\includegraphics[width=0.45\textwidth]{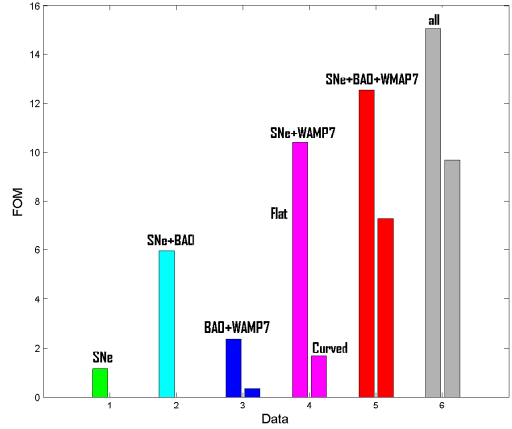}
\caption{FOM versus different data combinations. The left bars are for the flat CPL model and the right bars are for the curved CPL model.
For the curved CPL model, we use the following data sets: the combination of WMAP7 and BAO data, the combination of SNe and WMAP7 data, the combination of SNe, BAO and WMAP7 data,
and all data combined. For the flat CPL model, we use two more data sets: SNe data alone, and the combination of SNe Ia and BAO data.}
\label{fomfig}
\end{figure}

In this subsection, we apply the somewhat model-independent CPL parametrization \citep{cpl1,cpl2},
\begin{equation}
\label{lind}
w(z)=w_0+\frac{w_a z}{1+z},
\end{equation}
to test the effects of different combinations of data on constraining the property of dark energy.
For the flat CPL model, we have three model parameters $\mathbf{p}=(\Omega_{m},  \ w_0, \ w_a)$.
In Fig. \ref{figure3}, we show the marginalized $1\sigma$ and $2\sigma$ contour plots constrained from different combinations of data.
The FOM from different data is shown in Fig. \ref{fomfig}.
The $1\sigma$ uncertainties of the model parameters  are summarized in Table \ref{table2}.
Comparing the results from SNe Ia and BAO1 data, we see that BAO1 data give much better constraint on $\Omega_m$,
but their constraints on $w_0$ and $w_a$ are much worse. Therefore BAO data cannot constrain the dynamics of dark energy
although they give good constraints on $\Omega_m$ and the curved $\Lambda$CDM model.
When the BAO data are added to the SNe Ia data,
the uncertainty in $w_a$ is reduced more than half and the FOM becomes 5 times larger.
When WMAP7 data are added
to the SNe Ia data, the uncertainty in $w_a$ is reduced a little further and the FOM becomes almost 10 times larger.
Compared the results from SNe data alone with those from the combination of BAO and WMAP7 data, we see that
the SNe Ia data constrain better on $w_0$ and the combination of BAO and WMAP7 data constraints better on $w_a$;
the degeneracies among the model parameters from different data are different;
both BAO and WMAP7 data help reduce the uncertainties in $w_a$;
and the help from WMAP7 data is a little better.
The constraints on $w_0$ and $w_a$ from the combination of SNe and WMAP7 data are much better than those
from the combination of BAO and WMAP7 data, the FOM is almost 5 times larger. Note that we also fit two
nuisance parameters $\alpha$ and $\beta$ when we use SNe data, two nuisance parameters $\Omega_b h^2$ and $h$ when
we use BAO and WMAP7 data, four nuisance parameters $(\alpha,\beta,\Omega_b h^2, h)$ when we combine SNe data with BAO and WMAP7 data.
When the combined SNe Ia, BAO and WMAP7 data are used,
we get better constraints on $w_0$ and $w_a$.
The addition of $H(z)$ further reduces the uncertainties in $w_0$ and $w_a$. By using the constraints from the combination of all observational data,
we reconstruct $w(z)$ and $Om(z)$ and the results are shown in Fig. \ref{figure3}.
$\Lambda$CDM model is consistent with almost all the combinations of different data at the $1\sigma$ level.

\begin{table*}
\begin{center}
 \caption{The marginalized $1\sigma$ constraints on CPL model by different observational data.
 The top six rows are for the flat CPL model and the bottom four rows are for the curved CPL model.
 \label{table2}}
 \begin{tabular}{cccccc}
\hline Data & $\Omega_{m}$& $\Omega_{k}$& $w_0$& $w_a$ &FOM \\

\hline SNe & $0.31^{+0.09}_{-0.07}$ & & $-0.8^{+0.4}_{-0.2}$ &
$-3.1^{+2.4}_{-5.9}$  & $1.15$\\
\hline SNe+BAO & $0.28^{+0.03}_{-0.02}$ & & $-0.83^{+0.26}_{-0.19}$ &
$-2.11^{+1.27}_{-1.95}$  & $5.94$\\
\hline SNe+WMAP7 & $0.24^{+0.03}_{-0.02}$ & & $-0.9\pm 0.2$ &
$-1.1^{+0.8}_{-1.4}$  & $10.39$\\
\hline BAO+WMAP7 & $0.30\pm 0.04$ & & $-0.81^{+0.58}_{-0.59}$ & $-0.90^{+1.96}_{-1.91}$  & $2.38$\\
\hline SNe+BAO+WMAP7 & $0.28^{+0.02}_{-0.01}$ & &
$-1.12^{+0.27}_{-0.07}$ & $0.32^{+0.21}_{-1.63}$ & $12.52$\\
\hline All & $0.28^{+0.02}_{-0.01}$ & & $-1.00^{+0.17}_{-0.13}$ &
$-0.33^{+0.53}_{-1.03}$ & $15.0$\\ \hline
\hline BAO+WMAP7 & $0.35\pm 0.07$ & $-0.023\pm 0.013$ &
$0.4\pm 1.4$ & $-8.6^{+7.4}_{-7.2}$ & $0.36$ \\
\hline SNe+WMAP7 & $0.33^{+0.09}_{-0.06}$ & $-0.03^{+0.02}_{-0.03}$
& $-0.8^{+0.3}_{-0.2}$ & $-3.4^{+2.1}_{-5.0}$  & $1.67$\\
\hline SNe+BAO+WMAP7 & $0.28^{+0.02}_{-0.01}$ &
$-0.015^{+0.007}_{-0.008}$ & $-0.8\pm 0.2$ & $-2.02^{+1.35}_{-1.66}$ & $7.29$ \\
\hline All & $0.28^{+0.02}_{-0.01}$ & $-0.004^{+0.006}_{-0.007}$ &
$-0.97^{+0.27}_{-0.12}$ & $-0.57^{+0.65}_{-1.86}$ & $9.69$ \\
\hline
\end{tabular}
\end{center}
\end{table*}

\begin{figure*}
\includegraphics[width=0.95\textwidth]{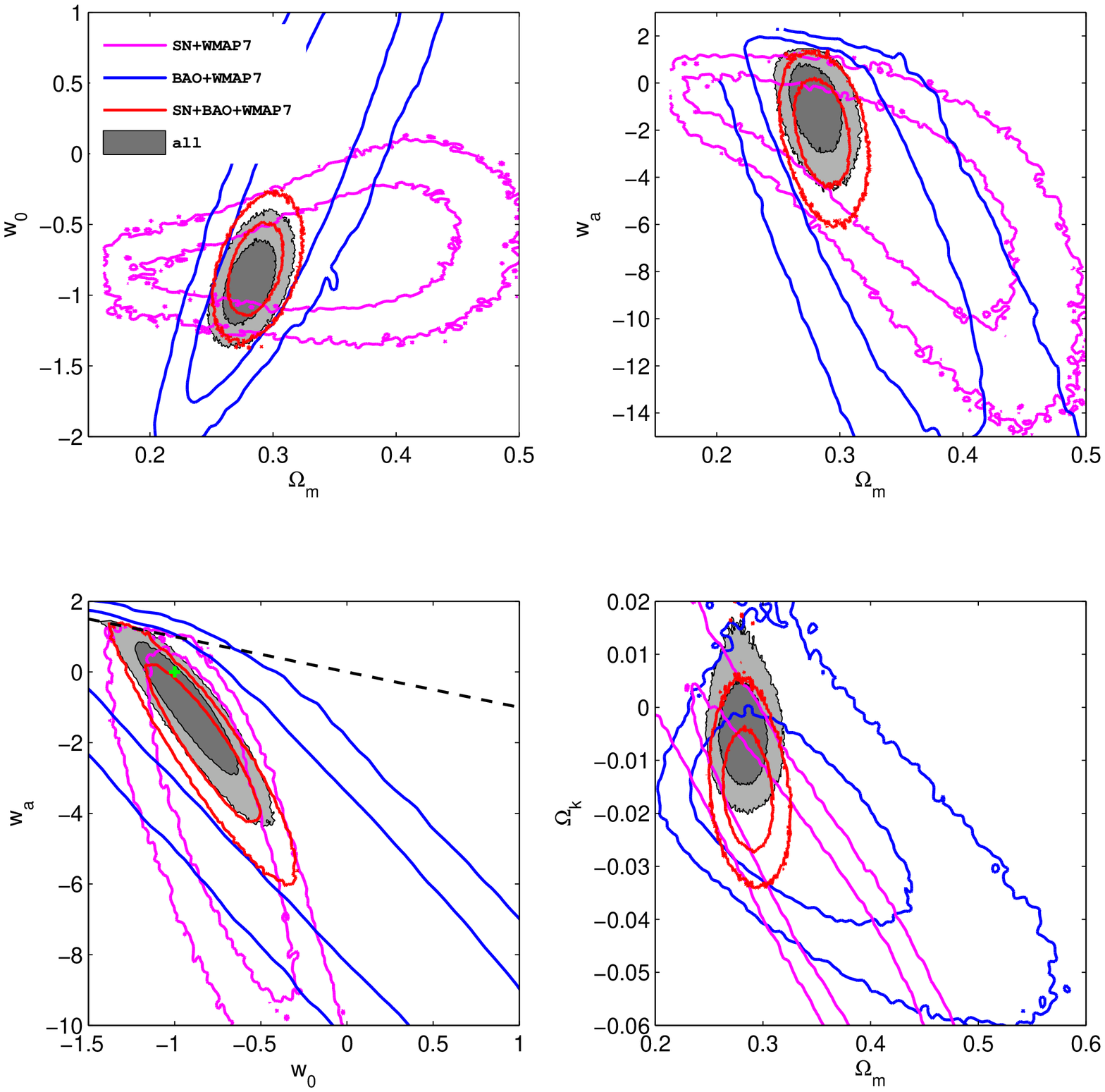}
\caption{The marginalized $1\sigma$ and $2\sigma$ contour plots for the curved CPL model.
the magenta lines label the constraints from the combination of SNe Ia and WMAP7 data,
the blue lines label the constraints from the combination of WMAP7 and BAO data,
the red lines label the constraints from the combination of SNe Ia, BAO and WMAP7 data,
and the shaded regions label the constraints from the combination of all the observational data.
The dashed line in
the $w_0$-$w_a$ contour denotes the condition $w_0+w_a=0$, and the + sign denotes the point corresponding to the $\Lambda$CDM model.}
\label{figure2}
\end{figure*}

\begin{figure*}
\includegraphics[width=0.95\textwidth]{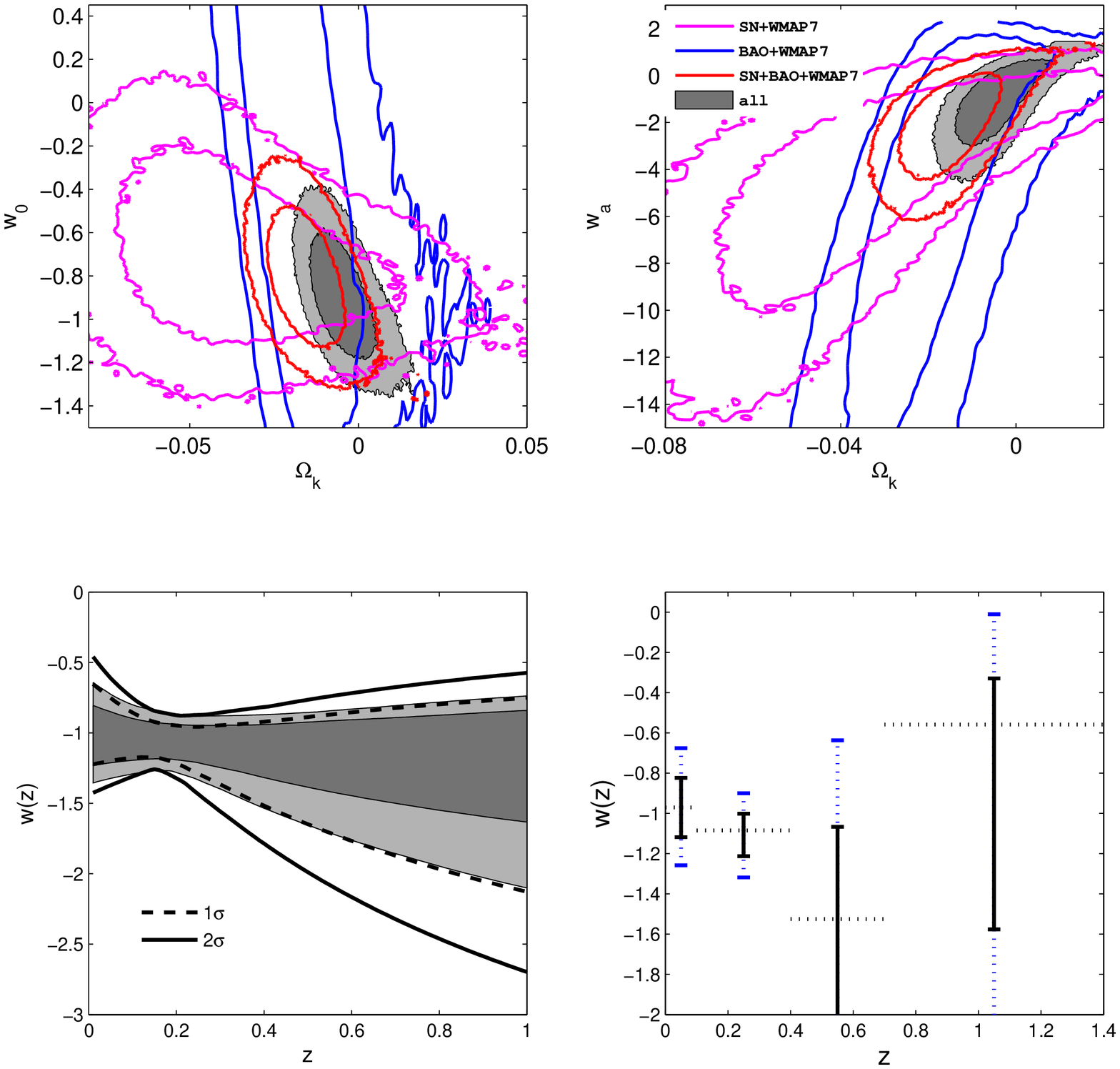}
\caption{The marginalized $1\sigma$ and $2\sigma$ constraints from observations.
In the upper panels,the magenta lines label the constraints from the combination of SNe Ia and WMAP7 data,
the blue lines label the constraints from the combination of WMAP7 and BAO data,
the red lines label the constraints from the combination of SNe Ia, BAO and WMAP7 data,
and the shaded regions label the constraints from the combination of all the observational data.
In the upper left panel,
we show the $\Omega_k$ and $w_0$ contours for the curved CPL model.
In the upper right panel, we show the $\Omega_k$ and $w_a$ contours for the curved CPL model.
In the lower left panel, we reconstruct
the evolution of $w(z)$ by using the constraints from the combination of all data for CPL model,
the shaded regions are for flat CPL model, and the black lines are
for curved CPL model. The dashed and solid lines are for the $1\sigma$ and $2\sigma$ uncertainties, respectively.
In the lower right panel, we show the observational constraints on $w(z)$ by using the piecewise parametrization,
the solid and dashed lines are for the $1\sigma$ and $2\sigma$ uncertainties, respectively.}
\label{figure4}
\end{figure*}

For the curved CPL model, we have four model parameters $\mathbf{p}=(\Omega_{m},\ \Omega_{k},  \ w_0, \ w_a)$,
and we use the following data:
the combination of SNe and WMAP7 data, the combination of BAO and WMAP7 data,
the combination of SNe, BAO and WMAP7, and all the data combined.
The $1\sigma$ and $2\sigma$ contours are shown in Fig. \ref{figure2} and the upper panels of Fig. \ref{figure4}.
The FOM from different data is shown in Fig. \ref{fomfig}.
The $1\sigma$ uncertainties of the parameters are summarized in Table \ref{table2}.
We see that the constraint on $\Omega_k$ from
the combination of BAO and WAMP7 data (the blues lines) is better than that from the combination of SNe Ia and WMAP7 data (the magenta lines).
The $\Omega_m$-$\Omega_k$ contour becomes much smaller when we combine SNe Ia, BAO and WMAP7 data (the red lines). The addition of $H(z)$ further reduces the errors
on $\Omega_k$ and moves the best-fitting value of $\Omega_k$ toward zero.
From the $w_0$-$w_a$ contours in Fig. \ref{figure2}, we see that the constraints
from the combination of SNe Ia and WMAP7 data are much better than those from the combination of BAO and WMAP7 data,
and the FOM is almost 5 times larger.
The uncertainties in $w_a$ from the combination of SNe Ia, BAO and WMAP7 data are reduced
more than half compared with those from the combination of SNe Ia and WMAP7 data, the FOM becomes
more than 4 times larger.
Although SNe Ia data do not provide tight constraints on $\Omega_m$ and $\Omega_k$,
their constraint on the equation of state parameter of dark energy is much better.
$\Lambda$CDM model (the cross) is outside the $1\sigma$ contour when we use the
combination of SNe Ia, BAO and WMAP7 data (the red lines). With the addition of $H(z)$ data,
the $w_0$-$w_a$ contour is further reduced and $\Lambda$CDM model is inside the $1\sigma$ contour.
By using the constraints from the combination of all observational data,
we reconstruct $w(z)$ and the result is shown in Fig. \ref{figure4}.

From the above discussion, we find that both BAO and WMAP7 data help SNe data tighten the constraint
on $\Omega_m$, hence better constrains the other model parameters for the flat case.
For the curved case, WMAP7 data help reduce the uncertainties in $\Omega_k$,
neither BAO nor WMAP7 data alone give good constraint on
$w_0$ and $w_a$, but both WMAP7 data and BAO data help SNe Ia data break the degeneracies
among the model parameters, hence tighten the constraint on the variation of
equation of state parameter $w_a$, and WMAP7 data do the job a little better.
SNLS3 SNe Ia data alone do not provide good constraints on $\Omega_m$ and $\Omega_k$,
but they provide good constrains on the parameters $w_0$ and $w_a$, especially on $w_0$,
so it is necessary to apply SNe Ia data
to probe the dynamical property of dark energy.
The addition of $H(z)$ data helps improve the constraints on the property of dark energy.
Due to the degeneracies among the model parameters,
we need to measure $\Omega_m$ and $\Omega_k$ more precisely in order to better probe the property of dark energy.
In other words, we need to combine different observational data such as SNe Ia, BAO, WMAP7 and $H(z)$ data as long as the
tensions among those data are not too big.

\subsection{Piecewise parametrization of $w(z)$}

Now we turn to probe the property of dark energy,
we apply all the observational data outlined in section 2 to the piecewise parametrization of $w(z)$ for flat case,
\begin{equation}
\label{wcez}
\Omega_{DE}(z)=(1-\Omega_{m})(1+z)^{3(1+w_N)}\prod_{i=1}^{N}(1+z_{i-1})^{3(w_{i-1}-w_i)},
\end{equation}
where $z_{i-1}\le z<z_i$, $z_0=0$, $z_1=0.1$, $z_2=0.4$, $z_3=0.7$ and $z_4=1.4$. We also assume that $w(z>1.4)=-1$.
Following \cite{huterer05}, we transform the parameters $w_i$
to the de-correlated parameters $\mathcal{W}_i$. The results of $\mathcal{W}_i$ are shown
in the lower right panel of Fig. \ref{figure4}.
The results are similar to those using Union2 SNe Ia data \citep{union2} and previous BAO data \citep{eg,wjp} in \cite{gong11},
and flat $\Lambda$CDM model
is consistent with this result.

\subsection{$q_1-q_2$ parametrization}

\begin{figure*}
\includegraphics[width=0.95\textwidth]{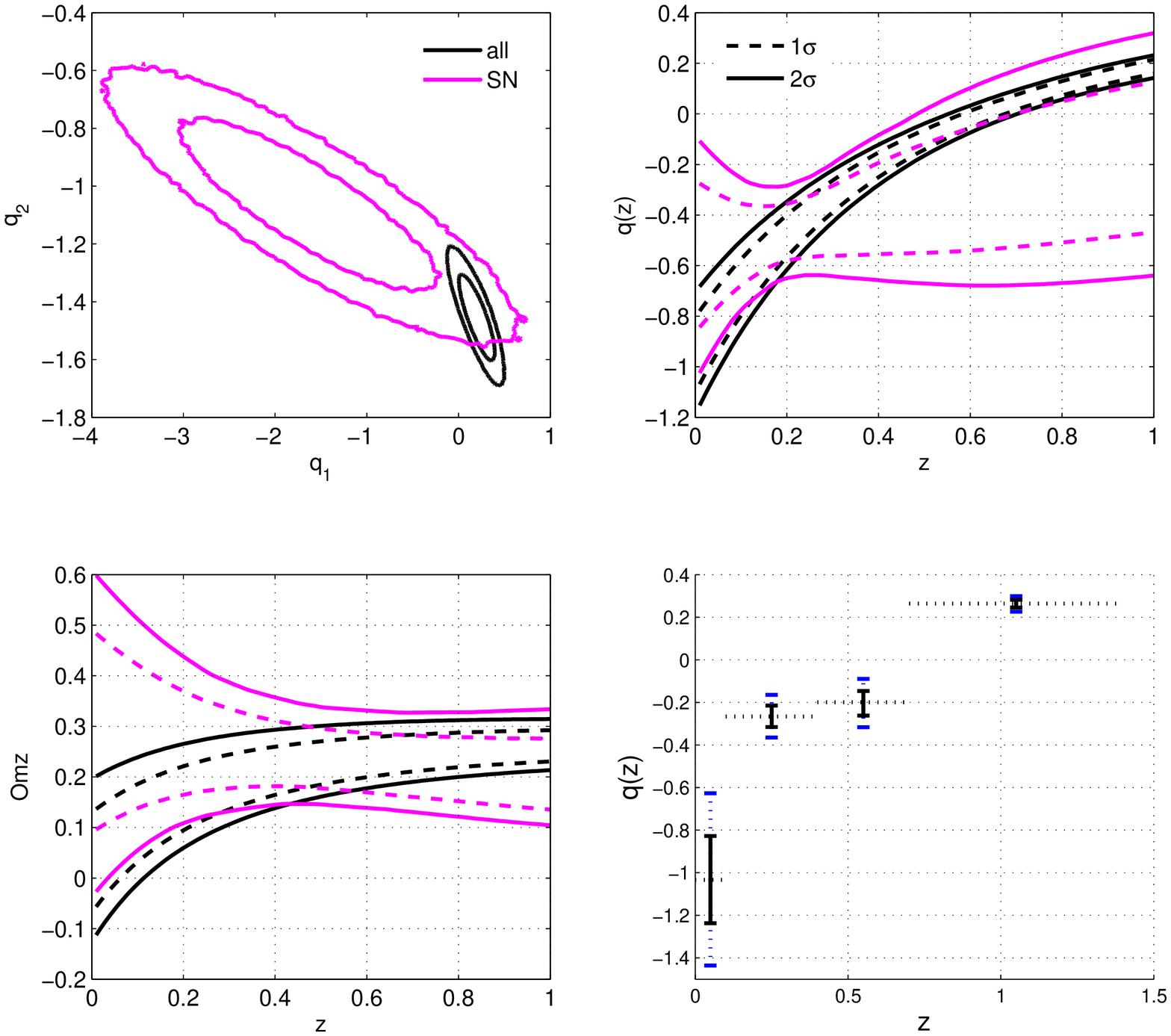}
\caption{The marginalized $1\sigma$ and $2\sigma$ constraints on
$q_1$-$q_2$ parametrization and the piecewise parametrization of $q(z)$,
the magenta lines represent the results obtained from SNe Ia data alone and the black lines represent the results
obtained from all the observational data.
The figures from upper left to lower right are labeled as (a)-(d), respectively. (a) shows the contour plots for $q_1$ and $q_2$,
(b) and (c) show the reconstruction of $q(z)$ and $Om(z)$, the dashed and solid lines in (b) and (c) represent the $1\sigma$ and
 $2\sigma$ uncertainties respectively, (d) shows the results for the piecewise parametrization of $q(z)$
constrained by all the observational data, the solid and dashed lines in (d) represent the $1\sigma$ and
 $2\sigma$ uncertainties respectively.}
\label{q12cont}
\end{figure*}

In this subsection, we reconstruct the deceleration parameter $q(z)$
with a simple two-parameter function \citep{gong07a},
\begin{equation}
\label{qmod2}
q(z)=\frac{1}{2}+\frac{q_1 z+q_2}{(1+z)^2}.
\end{equation}
This parametrization recovers the matter dominated epoch at high redshift
with $q(z)=1/2$. The dimensionless Hubble parameter is
\begin{eqnarray}
\label{hubsl2}
\begin{array}{ll}
E(z)&=\exp\left[\int_0^z[1+q(u)]d\ln(1+u)\right]\\
&=(1+z)^{3/2}\exp\left[\frac{q_2}{2}+\frac{q_1 z^2-q_2}{2(1+z)^2}\right].
\end{array}
\end{eqnarray}
Since $E^2(z)\approx (1+z)^3\exp(q_1+q_2)$ when $z\gg 1$, so the role of matter energy
density is played by the sum of the two parameters, $q_1+q_2=\ln\Omega_m$.
Although $\Omega_m$ and $\Omega_k$ are not model parameters in this parametrization,
the comoving distance depends on $\Omega_k$ through the function $S_k$,
in order to better constrain the model parameters $\mathbf{p}=(q_1, q_2)$,
we consider the flat case $\Omega_k=0$ only. As discussed above for the CPL model,
the flat assumption of $\Omega_k=0$ may impose biased prior in the estimation of cosmological
parameters due to the degeneracies among $\Omega_m$, $\Omega_k$ and
$w$ \citep{clarkson}. However, the only effect of $\Omega_k$
is through $S_k$, and $S_k(x)\approx x$ when $\Omega_k$ is small, so the impact of the
flat assumption is expected to be small. Fitting this model to SNe data alone,
we get the marginalized $1\sigma$ constraints $q_1=-1.68_{-0.87}^{+0.98}$ and $q_2=-1.06_{-0.2}^{+0.19}$.
The contour plot is shown in Fig. \ref{q12cont}(a). Using these results, we reconstruct $q(z)$ and $Om(z)$ and the results are shown in Figs. \ref{q12cont}(b)
and \ref{q12cont}(c). So $q(z)<0$ when $z\la 0.5$ at $2\sigma$ level and flat $\Lambda$CDM model is consistent
with the model at $1\sigma$ level. With the SNe Ia data alone, the evidence for current acceleration and past deceleration is very strong.

When we fit the model to BAO or WMAP7 data, we need to include the radiation-dominated era,
and the nuisance parameters $\Omega_b h^2$ and $\Omega_m h^2$ which are not appeared in the model
are just data fitting parameters,
so we do not apply BAO and WMAP7 data alone. For approximation, we
take the following Hubble parameter,
\begin{equation}
\label{hubsl3}
E^2(z)=\Omega_r(1+z)^4+(1+z)^3\exp\left[q_2+\frac{q_1 z^2-q_2}{(1+z)^2}\right],
\end{equation}
where the current radiation component $\Omega_{r}=4.1736\times 10^{-5}h^{-2}$ \citep{wmap7}.
Fitting the model to the combined SNe Ia, BAO, WMAP7 and $H(z)$ data,
we get the marginalized $1\sigma$
constraints, $q_1=0.20\pm 0.13$ and $q_2=-1.45\pm 0.1$. The contour plot is shown in Fig. \ref{q12cont}(a).
Compared this result with that obtained from SNe Ia data alone, we find that
they are inconsistent at $1\sigma$ model, this shows the tension between SNe Ia, BAO and WMAP7 data
in fitting this model. Using the $q_1$-$q_2$ contour, we reconstruct $q(z)$ and $Om(z)$
and the results are shown in Figs. \ref{q12cont}(b)
and \ref{q12cont}(c). We find that $q(z)$ increases with the redshift and $q(z)< 0$ when $z\la 0.5$ at $2\sigma$ level,
the flat $\Lambda$CDM model is inconsistent with the model at $2\sigma$ level. These results may suggest that
the approximation (\ref{hubsl3}) is not good at high redshift. Since $\Omega_m$ does not appear in this model,
the application of BAO and WMAP7 data may not be straightforward, this needs to be further studied.

\subsection{Piecewise parametrization of $q(z)$}

We also apply the piecewise parametrization to study the property of the deceleration parameter $q(z)$.
For $z_{i-1}\le z<z_i$, we have
\begin{equation}
\label{qzcez}
E(z)=(1+z)^{1+q_N}\prod_{i=1}^{N}(1+z_{i-1})^{q_{i-1}-q_i}.
\end{equation}
In this model, we have four parameters $\mathbf{p}=(q_1,\ q_2,\ q_3,\ q_4)$. We think this
model approximates the behavior of $E(z)$ in the redshift range $z\la 1.5$. In the radiation
dominated era, we add the radiation contribution also.
Again we follow \cite{huterer05} to transform the correlated parameters $q_i$ to uncorrelated  ones.
Fitting the model to all observational data, we reconstruct the evolution of $q(z)$
and the results are shown in Fig. \ref{q12cont}(d). Similar to that obtained by Union2 SNe Ia data \citep{gong11},
we find that $q(z)<0$ when $z\la 0.6$ and $q(z)>0$ at high redshift, so the
evidences for current acceleration and past deceleration are very strong,
and the transition redshift is around $z_t\sim 0.7$.

\section{Conclusions}

It is well known that BAO data are more sensitive to $\Omega_m$ and WMAP7 data are more sensitive to $\Omega_k$.
As more data points become available and the data become more accurate,
we are able to constrain the cosmological parameters better.
The constraints on $\Lambda$CDM model from SNe Ia data alone are similar to those
from BAO data alone as shown in Figs. \ref{lcdmcont} and \ref{baoclmcont},
the constraints on $\Omega_m$ by BAO data and the constraints on $\Omega_m$ and $\Omega_k$ by BAO1 data are even much better than those by SNe Ia data alone,
and the addition of SNe data to the combination of BAO and WMAP7 data has little effect in improving
the constraints of $\Omega_m$ and $\Omega_k$.
Applying BAO data to MHDE and CPL models,
we find that the constraints on the dynamical behaviour of dark energy
from BAO data alone are much worse than those from SNe Ia data alone. Although
SNe Ia data alone are not able to provide good constraints on $\Omega_m$ and $\Omega_k$, they provide much better constraint
on the equation of state parameter of dark energy compared with that from BAO and WMAP7 data alone.
Since the way that the model parameters are degenerated is different for SNe Ia, BAO and WMAP7 data alone, it
is necessary to combine different data sets to get better constraint on the property of dark energy.

For the flat MHDE model, as shown in Fig. \ref{figure5}(a), although BAO1 data provide good constraint
on $\Omega_m$, but the constraints on $N$ are much worse, this suggests that BAO data alone are not able to constrain
the dynamics of dark energy. Both WMAP7 and BAO data help SNe data greatly improve the constraint on $\Omega_m$, and
the help from WMAP7 is even a little greater. The constraints on $\Omega_m$ and $N$ from the combination of
BAO and WMAP7 are similar to those from the combination of SNe and WMAP7. The addition of $H(z)$
data to the combination of SNe Ia, BAO and WMAP7 data has little effect on improving the constraints.
For the curved MHDE model, as shown in Figs. \ref{figure5}(b)-(d), the combination of BAO and WMAP7 data gives more stringent constraints on $\Omega_m$ and $\Omega_k$
than the combination of SNe and WMAP7 data does, but the constraints on $N$ from both combinations are similar.
The addition of $H(z)$ data helps tighten the constraints a little further.
We also find that $\Lambda$CDM model is favoured against DGP model.

For the flat CPL model, as shown in Fig. \ref{figure3}, although BAO1 data provide good
constraint on $\Omega_m$, but the constraints on $w_0$ and $w_a$ are much worse, this confirms that BAO data alone are not able to constrain
the dynamics of dark energy.
We get similar constraints on $w_0$ and $w_a$ for SNe Ia data alone and the
combination of BAO and WMAP7 data, although the constraints on $\Omega_m$ from the latter combination are much better.
When the BAO data are added to the SNe Ia data,
the uncertainty in $w_a$ is reduced more than half and the FOM becomes 5 times larger as shown in Table \ref{table2} and Fig. \ref{fomfig}.
When WMAP7 data are added
to the SNe Ia data, the uncertainty in $w_a$ is reduced almost 4 times and
the FOM becomes almost 10 times larger as shown in Table \ref{table2} and Fig. \ref{fomfig}.
Both BAO and WMAP7 data help SNe Ia data reduce the uncertainties in $w_a$,
and the help from WMAP7 data is a little better.
The addition of $H(z)$ data to the combination of SNe Ia, BAO and WMAP7 data has little effect on improving the constraints.
$\Lambda$CDM model is consistent with all the observational data.
This point is further supported by the reconstruction of $w(z)$ and $Om(z)$ as shown in Figs. \ref{figure3} and \ref{figure4}.

For the curved CPL model, as shown in Figs. \ref{figure2} and \ref{figure4}, we find that the combination of BAO and WMAP7
data gives better constraints on $\Omega_m$ and $\Omega_k$
than those from the combination of SNe Ia and WMAP7, but the constraints on $w_0$ and $w_a$ from the first combination
are much worse than those from the latter combination.
The $1\sigma$ uncertainties of $w_0$ and $w_a$ from the latter combination
are greatly reduced and the FOM becomes 5 times larger as shown in Table \ref{table2} and Fig. \ref{fomfig}.
The $1\sigma$ uncertainty of $w_a$ is reduced
more than half when we add BAO data to the combination of SNe Ia and WMAP7 data,
and the FOM becomes more than 4 times larger as shown in Table \ref{table2} and Fig. \ref{fomfig}.
The $1\sigma$ uncertainty of $w_a$
is reduced more than 5 times when we add SNe Ia data to the combination of BAO and WMAP7 data,
and the FOM becomes more than 20 times larger as shown in Table \ref{table2} and Fig. \ref{fomfig}.
$H(z)$ data help move the best-fitting value of $\Omega_k$ towards zero
and make the model more compatible with $\Lambda$CDM model.
This effect of $H(z)$ which moves the best-fitting value of $\Omega_k$ towards zero
for the curved $\Lambda$CDM, the curved MHDE and the curved CPL model seems to be general,
this needs to be further studied.

To study the acceleration of the expansion of the Universe, we reconstruct the deceleration parameter $q(z)$
with a simple two-parameter function and the piecewise parametrization which approximates the evolution of the Universe
in the redshift $z\la 1.5$. For the SNe Ia data only, we see strong evidence that $q(z)<0$ in the redshift $z\la 0.5$
as shown in Fig. \ref{q12cont}.
The $1\sigma$ contour from SNe Ia data only is inconsistent with that from the combination of all data,
it seems that there exists some tensions between SNe Ia data and other data;
however, the inconsistency may come from the way we apply the BAO and WMAP7 data. Note that the nuisance model
parameters $\Omega_m h^2$ and $\Omega_b h^2$ do not appear in the $q(z)$ parameterizations, but BAO and WMAP7 data
depend on those parameters, so we must be careful of applying those nuisance parameters, this needs to be further studied.

For the curved $\Lambda$CDM model, BAO1 data alone give much better constraint than SNe Ia data alone do,
the combination of BAO and WMAP7 data constrains $\Omega_m$ and $\Omega_k$
much better than the combination of SNe and WMAP7 data.
For the MHDE model, the constraints on the only dark energy
parameter $N$ from BAO or BAO1 data are much worse than those from SNe Ia data
although the former data give much better constraints on $\Omega_m$, the constraints on $N$ from
the combination of BAO and WMAP7 data are similar to those from the combination of SNe Ia and WMAP7 data,
although the constraints on $\Omega_m$ and $\Omega_k$ from the former data are much better for the curved MHDE model.
For the CPL model, we have two dark energy parameters $w_0$ and $w_a$. SNe Ia data alone
give much better constraints on $w_0$ and $w_a$ than BAO1 data alone do. The
combination of SNe Ia and WMAP7 data give more stringent
contours of $w_0$ and $w_a$ than the combination of BAO and WMAP7 data.

For the flat models, both BAO and WMAP7 data help SNe Ia data greatly improve the constraint on $\Omega_m$
and the help from WMAP7 is a little better, therefore both BAO and WMAP7 data help SNe Ia data tighten
the constraint on the property of dark energy. The addition of $H(z)$
data to the combination of SNe Ia, BAO and WMAP7 data has little effect on improving the results.
Although BAO and WMAP7 data provide reasonably good constraints on $\Omega_m$ and $\Omega_k$,
they are not able to constrain the dynamics of dark energy,
we need SNe Ia data to probe the property of dark energy, especially
the variation of the equation of state parameter of dark energy $w_a$. This point was well known due to the different directions of
the degeneracy obtained from different data,
here we confirm the point with the updated BAO data which constrain the $\Lambda$CDM model even better
than SNe Ia data do.
The addition of BAO data helps reduce the error on $\Omega_m$ and the addition
of CMB data helps reduce the error on $\Omega_k$, so both BAO and CMB data
help SNe Ia data tighten the constraints on the property of dark energy due to degeneracies
among $\Omega_m$, $\Omega_k$ and $w(z)$, but neither data alone can be used to probe
the dynamical property of dark energy.
For the SNLS SNe Ia data, the nuisance parameters $\alpha$ and $\beta$ are consistent for all different combinations
of data. Their impacts on the fitting of cosmological parameters are minimal.

\section*{acknowledgments}

This work was partially supported by
the National Basic Science Programme (Project 973) of China under
grant Nos. 2010CB833004 and 2012CB821804, the NNSF of China under grant Nos. 10935013 and 11175270,
the Project of Knowledge Innovation Program (PKIP) of Chinese Academy of Sciences, Grant No. KJCX2.YW.W10,
and the Fundamental Research Funds for the Central Universities.

%\bibliographystyle{mn2e}
%\bibliography{deobs} % all references

\end{document}